\documentclass[aip,jcp,reprint,floatfix,superscriptaddress]{revtex4-1}

\usepackage{graphicx,textcomp,hyperref,natbib,color,amsmath}

\hypersetup{
  colorlinks   = true, 
  urlcolor     = blue, 
  linkcolor    = black, 
  citecolor   = blue 
}

\begin{document}
\title{Photodissociation spectroscopy of the dysprosium monochloride molecular ion}
\author{Alexander Dunning}
\email{alexander.dunning@gmail.com}
\affiliation{Department of Physics \& Astronomy, University of California, Los Angeles, CA 90095, USA}

\author{Alexander Petrov}
\altaffiliation{Alternative address: NRC `Kurchatov Institute' PNPI 188300, Division of Quantum Mechanics, St. Peters-burg State University, 198904, Russia}
\affiliation{Department of Physics, Temple University, Philadelphia, PA 19122, USA}

\author{Steven J. Schowalter}
\author{Prateek Puri}
\affiliation{Department of Physics \& Astronomy, University of California, Los Angeles, CA 90095, USA}
\author{Svetlana Kotochigova}
\affiliation{Department of Physics, Temple University, Philadelphia, PA 19122, USA}
\author{Eric R. Hudson}
\affiliation{Department of Physics \& Astronomy, University of California, Los Angeles, CA 90095, USA}

\begin{abstract}
We have performed a combined experimental and theoretical study of the photodissociation cross section of the molecular ion DyCl$^+$. The photodissociation cross section for the photon energy range 35,500 cm$^{-1}$ to 47,500 cm$^{-1}$ is measured using an integrated ion trap and time-of-flight mass spectrometer; we observe a broad, asymmetric profile that is peaked near 43,000 cm$^{-1}$. The theoretical cross section is determined from electronic potentials and transition dipole moments calculated using the relativistic configuration-interaction valence-bond and coupled-cluster methods. The electronic structure of DyCl$^+$ is extremely complex due to the presence of multiple open electronic shells, including the 4f$^{10}$ configuration. The molecule has nine attractive potentials with ionically-bonded electrons and 99 repulsive potentials dissociating to a ground state Dy$^+$ ion and Cl atom. We explain the lack of symmetry in the cross section as due to  multiple contributions from one-electron-dominated transitions between the vibrational ground state and several resolved repulsive excited states. 
\end{abstract}
\maketitle

\section{Introduction}

Molecular ions provide a platform for many advances in quantum physics and chemistry \cite{Carr2009}. They offer the same rich rovibrational structure and internal fields as their neutral counterparts, yet they are more straightforward to trap on long timescales, and they offer the potential for improved quantum coherence \cite{Schuster2011}. These advantages have motivated many recent research efforts centered around molecular ions, including studies of the time-variation of fundamental constants \cite{Pasteka2015a}, searches for the electric dipole moment of the electron \cite{Meyer2006,Leanhardt2011}, and designs of quantum computing architectures in which molecular ion qubits are addressed by microwave \cite{Schuster2011} and laser \cite{Yun2013} fields. Necessarily, much progress has been made toward cooling molecular ions to quantum degeneracy: millikelvin translational temperatures and Coulomb crystallisation have been demonstrated with hydride-like \cite{Molhave2000} and more complex \cite{Ostendorf2006} molecular ions via sympathetic cooling of the molecular species by co-trapped laser-cooled atomic ions; and quenching of the rotational and vibrational motion of molecular ions has been achieved through sympathetic cooling by a cryogenic buffer gas \cite{Hansen2014}, and by a co-trapped cloud of ultracold neutral atoms \cite{Rellergert2013}, respectively.

However, since the field of ultracold molecular ion physics is relatively young, one significant obstacle to progress in these experiments is the small pool of available spectrosopic data. In the past three decades, there have been relatively few experimental spectroscopic studies of diatomic molecular ions (see Saykally and Woods' summary of pre-1980 work in Ref. \onlinecite{Saykally1981}, and as a representative yet non-exhaustive list of the work since then, see Refs.~\onlinecite{Shenyavskaya1980,Balfour1980,Ramsay1982,Merer1984,Chanda1995,Wuest2004,Goncharov2006,Merritt2009,Chen2011,Antonov2013}).

In this report we present spectroscopic data, and a detailed theoretical discussion, of the molecular ion DyCl$^+$, which is an interesting candidate for molecular laser cooling, ultracold chemistry in atom-ion systems, and in the long-term, scalable quantum computing. The interest in DyCl$^+$, and indeed the lanthanide halides in general, stems from the notion that unpaired electrons in the $4f$ core of the lanthanide element play little or no part in the bonding of the molecular system \cite{Lanza2008}. Since excited states within the $4f$ levels are often present at optical intervals, and Laporte-forbidden $f-f$ transitions may be activated by vibrations and interactions with the ligand\cite{Hatanaka2014}, this could give rise to optically addressable transitions with highly diagonal Franck-Condon factors.

Owing to a heightened complexity, theoretical approximations regarding core electron configurations and molecular potentials in the lanthanide halides remain contestable. Demonstrating agreement between experiment and theory will naturally help to strengthen the validity of our chosen theoretical methods.

The remainder of this manuscript is structured as follows. In Sec.~\ref{TheorySec}, we describe the theoretical calculations of the DyCl$^+$ molecular potentials, along with predictions of transition strengths between bound and repulsive states. In Sec.~\ref{ExptSec}, we outline the experimental methods for measuring the photodissociation cross section. We present a comparison of our theoretical and experimental results in Sec.~\ref{ResultSec}, along with some discussion regarding bound excited state potentials in the molecular ion, before summarising in Sec.~\ref{SummarySec}.

\section{Theoretical Calculations}\label{TheorySec}

The DyCl$^+$ molecule is a truly relativistic molecule with an intricate electronic structure, whose complexity is  due to a partially-filled Dy$^+$ anisotropic inner $4f^{10}$ shell, which lies beneath an open $6s$ shell. In fact, the ground state of the Dy$^+$ ion has a very large total angular momentum $j=17/2$ (excluding nuclear spin) indicative of strong alignment of its electron spins. Coupling the Dy$^+$ ion with the Cl atom, which has an open $3p^5$ shell, makes the molecular structure even more complex. 

In our dissociation experiment with photon energies around 40,000 cm$^{-1}$, we need only consider transitions 
between relativistic potentials that dissociate to the energetically-lowest four  limits of Dy$^+(4f^{10}6s)$ + Cl$(3p^5)$. The atomic states and energies of these limits are given in Table~\ref{states}. The angular momentum coupling of the DyCl$^+$ molecule is described by the Hund's case (c) coupling scheme and labeled by  $\Omega$, the absolute value of the projection of the total electronic angular momentum $\vec j =\vec \jmath_1 + \vec \jmath_2$ on the intermolecular axis. Here $\vec \jmath_1$ and $\vec \jmath_2$ are the total atomic angular momenta of Dy$^+$ and Cl, respectively. Table ~\ref{states} shows the number of potentials for each $\Omega$ yielding a total of 108 potentials.

\begin{table}
\begin{tabular}{c@{$\quad$}c @{\hskip 0.5cm} c @{\hskip 0.5cm} ccccccccccc}
\toprule
     &     &   $E/(hc)$   & \multicolumn{11}{c}{$\Omega$}\\
\cline{4-14}   $j_1$   & $j_2$ &  (cm$^{-1}$) & 10 & 9 & 8 & 7 & 6 & 5 & 4 & 3 & 2 & 1 & 0 \\
\colrule
17/2 & 3/2 &  0.0     &   1 & 2 & 3 & 4 & 4 & 4 & 4 & 4 & 4 & 4 & 4       \\
15/2 & 3/2 &  828.314   &     &1  & 2 & 3 & 4 & 4 & 4 & 4 & 4 & 4 & 4   \\
17/2 & 1/2 &  882.3515  &     & 1 & 2 & 2 & 2 & 2 & 2 & 2 & 2 & 2 & 2 \\
15/2 & 1/2 &  1710.6655  &     &   & 1 & 2 & 2 & 2 & 2 & 2 & 2 & 2 & 2 \\
\botrule
\end{tabular}
\caption{The number of relativistic Hund's case (c) potentials with label $\Omega$ for the energetically-lowest  four dissociation 
limits of Dy$^+$$[4f^{10}(^5{\rm I}_8)6s_{1/2}\,(8,1/2)_{j1}]$+ Cl$[3p^5(^2{\rm P}_{j2})]$, where $j_1$ and $j_2$ are
the total angular momentum quantum numbers of Dy$^+$ and Cl, respectively [The `$(8,1/2)$' notation indicates the core and outer total electronic angular momenta in the j-j coupling regime]. Here $h$ is Planck's constant and $c$ is the speed of light.}
\label{states}
\end{table}

We have determined the DyCl$^+$ potentials, for the bound and repulsive states corresponding to the configurations given in Table~\ref{states}, as a function of internuclear separation $R$, based on a two-step approach. First, we calculate relativistic potentials  with a relativistic configuration-interaction valence-bond (RCI-VB) method \cite{Kotochigova2005} using a relatively small basis set limited by computational resources. In the RCI-VB method the molecular wave function is constructed from atomic Slater determinants with localized one-electron orbitals found by numerically solving the Dirac-Fock (DF) equation for occupied orbitals and  DF or Sturmian equations for virtual or unoccupied orbitals. We perform all-electron calculations where, in principle, the dynamics of  each electron is accounted for. To restrict the size of the Hamiltonian, excitations from the closed shells of Dy$^+$ are not allowed, while excitations of electrons in the open 4f$^{10}$ and 6s shells and into unoccupied 6p and 5d virtual shells are allowed. These four shells form the active space of Dy$^+$. Similarly, for the Cl atom, shells up to 2p remain closed while excitations of the occupied 3s and 3p shells into the virtual 3d and 4s are included.

Our calculations show that there are nine ionically-bonded attractive potentials, one for each $\Omega=0$ up to 8, with splittings that are much less than  their depth. The $\Omega=8$ potential is the deepest. The remaining 99 potentials are repulsive or barely attractive. We denote the potentials by $V_{n,\Omega}^{\rm VB}(R)$, where $n=1,2,3\dots$ labels the $\Omega$ potentials with increasing energy, and uniformly shift all potentials such that the energetically-lowest potential approaches zero energy in the limit of large separation.

The structure of the molecular potentials can be qualitatively understood by
noting that when the $6s$ valence electron of Dy$^+$ is transferred to
Cl, the ionically-bonded and thus the deeply-bound molecule
Dy$^{++}$Cl$^-$ is formed. In this molecule, the Cl$^-$ has a closed $3p^6$
shell, while Dy$^{++}$ only has one open electron shell, the $4f^{10}$, with
total angular momentum $j=8$. Consequently, we have nine deep potentials,
one for each  projection quantum number $\Omega=0$ to $8$.

Next, we improve the nine attractive potentials with the help of  {\it non-relativistic} coupled-cluster calculations, where we can include more electron-electron correlations with much larger basis sets as long as a single determinant dominates the bond, and with the realization that (small) splittings between potentials converge more rapidly than the absolute binding energy. We, therefore, compute the only non-relativistic potential consistent with $\Omega=8$, corresponding to the Dy$^{++}$ ``stretched-state'' with total electron spin $S=2$ and projection of the
electron orbital angular momentum $\Lambda=6$ using single, double, and perturbative triple excitations (CCSD(T)).  The coupled-cluster method does not converge for $\Omega<8$  as then multiple determinants control the bonding.

For Dy we use the scalar relativistic Stuttgart ECP28MWB pseudopotential and
associated atomic basis sets (14s13p10d8f6g)/[10s8p5d4f3g] \cite{Lim2005}.
The basis set def2-QZVPP (20s14p4d2f1g)/[9s6p4d2f1g] is used for Cl
\cite{Weigend2005}. The core electrons 1s$^2$2s$^2$2p$^6$ for Cl  and
4s$^2$4p$^6$4d$^{10}$ for Dy are not correlated, leaving core-valence electrons 
of 3s$^2$3p$^5$ in Cl and 5s$^2$5p$^6$4f$^{10}$ in Dy correlated. We find that the potential,
$V^{\rm CC}(R)$, has a minimum at $R_e = 4.63a_0$ with depth $D_e /(hc)
=32508$ cm$^{-1}$ and where $a_0$ is the Bohr radius. For
$^{161}$Dy$^{35}$Cl$^+$ it has a $\omega_e$/(hc) = 441 cm$^{-1}$ and  120
bound states. Unsurprisingly, we find that $V^{\rm CC}(R)$ is deeper than
any of the relativistic potentials. Again this potential is shifted such
that it approaches zero at large $R$. Finally, we construct $V_{n,\Omega}(R)
= V^{\rm CC}(R) + (V_{n,\Omega}^{\rm VB}(R)-V_{1,8}^{\rm VB}(R))$ for the
nine attractive $\Omega$ potentials. The potential-energy functions for
remaining mainly-repulsive electronic states are taken from the RCI-VB
calculation.

\begin{figure}
\includegraphics[width=8.2cm]{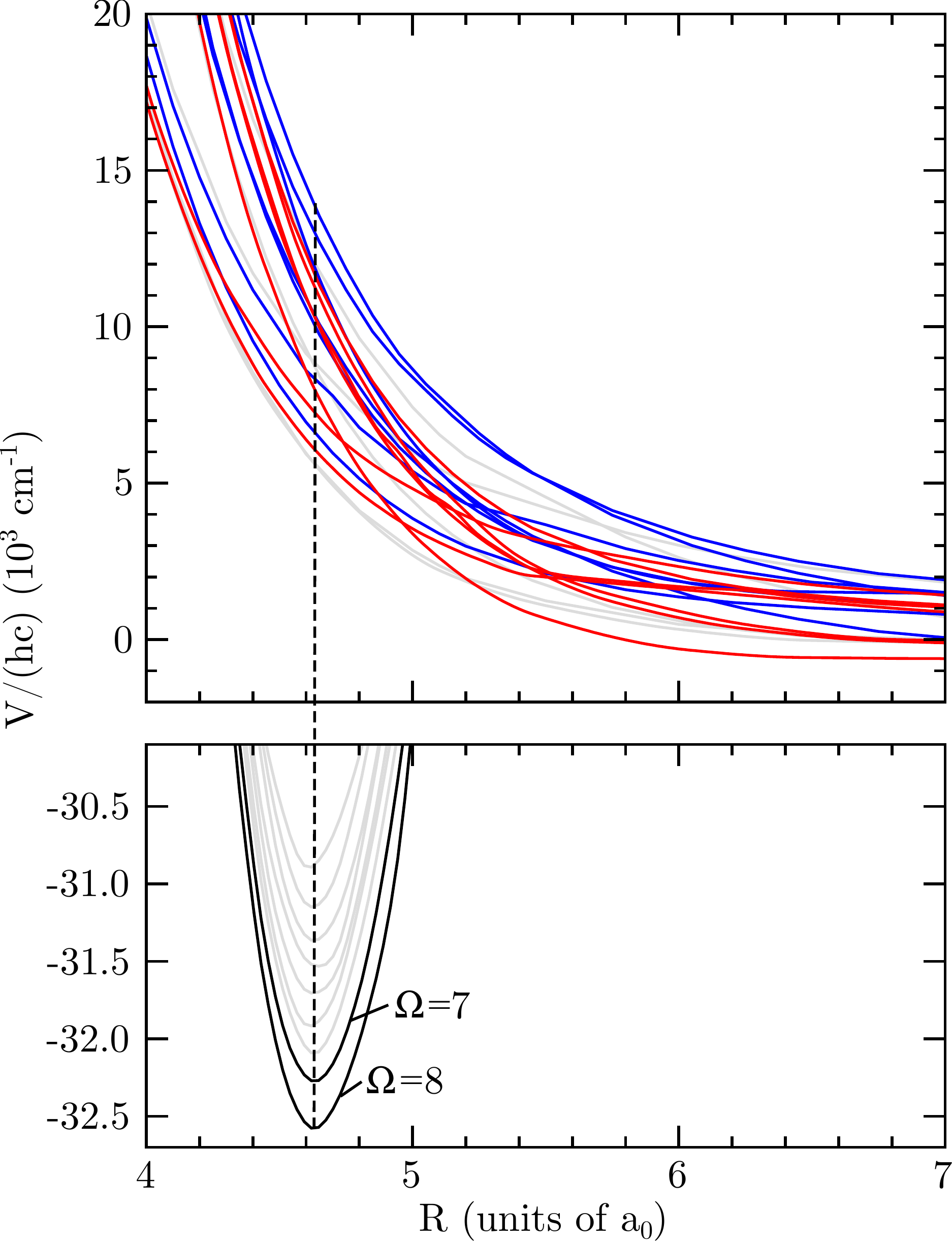}
\caption{(color online) Potential energy curves of the DyCl$^+$ molecule as a function of $R$. 
The bottom panel shows the nine attractive, deeply-bound potentials, one for each $\Omega$  
from 0 to 8. The two black curves correspond to the  $\Omega=$8 and 7 potentials 
populated in our photodissociation experiment. Other potentials are shown by grey lines.  
The top panel shows repulsive and weakly-bound  $\Omega'$=6, 7, 8, and 9 potentials. The seven red 
curves are $\Omega'$=7, 8, and 9 potentials with a large transition dipole moment to the $\Omega=8$ ground state and dissociate to Dy$^+$+Cl limits. The other seven blue curves are $\Omega'$=6, 7, and 8 potentials that
have a large transition dipole moments with the ground $\Omega=7$ state. Other potentials are shown with grey lines. 
The dashed black line indicates vertical transitions  from the $v=0$ rovibrational level of 
the $\Omega=8$ potential to  repulsive potentials driven by one-color laser radiation.
}
\label{curves}
\end{figure}

The minima of the nine adjusted attractive potentials as well as the repulsive potentials relevant for the simulation of our photodissociation experiment are shown in Fig.~\ref{curves}.   In addition, Table~\ref{bound} shows the first thirty $J=8$ vibrational energies of the ground state $n,\Omega=1,8$ or X  potential for two isotopes of DyCl$^+$. 

\begin{table}[b] 
\begin{tabular}{c cc c cc}
\toprule & \multicolumn{2}{c}{$E_v/(hc)$ (cm$^{-1}$)}  &  & \multicolumn{2}{c}{$E_v/(hc)$ (cm$^{-1}$)} \vspace{0.06cm}\\
\cline{2-3}\cline{5-6} $v$ & $^{161}$Dy$^{35}$Cl$^+$ & $^{161}$Dy$^{37}$Cl$^+$
                     & $v$ & $^{161}$Dy$^{35}$Cl$^+$ & $^{161}$Dy$^{37}$Cl$^+$\\
\colrule
0   &   -32351   &-32356      &15 &   -25854      &-26413\\
1   &   -31914   &-31928      &16 &   -25435      &-25590\\
2   &   -31495   &-31518      &17 &   -25019      &-25182\\
3   &   -31077   &-31110      &18 &   -24605      &-24777\\
4   &   -30648   &-30692      &19 &   -24192      &-24373\\
5   &   -30205   &-30260      &20 &   -23780      &-23970\\
6   &   -29752   &-29818      &21 &    -23367     &-23566\\
7   &   -29295   &-29372      &22 &    -22955     &-23163\\
8   &   -28841   &-28926      &23 &    -22543     &-22760\\
9   &   -28398   &-28491      &24 &    -22133     &-22359\\
10  &   -27968   &-28069      &25 &    -21725     &-21958\\
11  &   -27543   &-27652      &26 &     -21319    &-21560\\
12  &   -27119   &-27238      &27 &     -20916    &-21164\\
13  &   -26697   &-26825      &28 &     -20516    &-20772\\
14  &   -26275   &-26413      &29 &     -20120    &-20382\\
\botrule
\end{tabular}
\caption{The predicted binding energy of the first thirty vibrational energies of the lowest $n,\Omega=1,8$ or X  potential of the  $^{161}$Dy$^{35}$Cl$^+$ and $^{161}$Dy$^{37}$Cl$^+$ isotopes in the $J=8$ rotational state. }
\label{bound}
\end{table} 

\begin{table}[t]
\caption{$^{161}Dy^{35}Cl^+$ molecular spectroscopic constants for the deepest $\Omega$ = 7 and 8 potentials} 
\begin{tabular}{lcccc}
\toprule
State          &  $R_e$    &  $D_e$       & $\omega_e$ & $B_e$   \\ 
                     & [$a_0$] & [cm$^{-1}$] & [cm$^{-1}$]   & [cm$^{-1}$] \\
\hline 
$\Omega$=8  & 4.63   &  32~508     &  441              & 0.0977 \\
$\Omega$=7  & 4.63   &  32~273     &  466              & 0.0977\\
\botrule
\end{tabular}
\label{constants}
\end{table}

In our experiment, we estimate the internal blackbody redistribution rate to be around 1~Hz, and so we predict that the molecules are approximately in rovibronic thermal equilibrium with the surroundings when photodissociation begins after 1.7~s of trapping. This approximation is validated by previous studies of molecular ions with more straightforward electronic structure using the same apparatus \cite{Chen2011, Puri2014, Rellergert2013}, where 300~K-based calculations were found to match experimental data well.

For simplicity we assume that the photodissociation cross section is independent of $J$ and only thermalize  the vibrational distribution. Assuming thermally-populated levels at $T=300$~K, 81.6\% and 10\% of the population is in the $v$=0 and $v$=1 vibrational levels of the ground $\Omega$=8 potential, respectively. The only remaining level which is non-negligibly populated in thermal equilibrium is the $v$=0 vibrational level of the ground $\Omega$=7 potential, so we approximate all of the remaining 8.4\% to be here in the calculation.
The spectroscopic constants for the ground $\Omega$=8 and 7 potentials are given in Table~\ref{constants}.

The Franck-Condon principle states that the internuclear separation remains approximately unchanged during an electronic transition \cite{Herzberg1957}. In Fig.~\ref{curves} this principle is illustrated by a vertical line at the peak internuclear separation of the $v=0$ state, and employing the reflection approximation \cite{Lefebvre-Brion2004}, we predict that photon energies between 37,500 cm$^{-1}$ and 44,500 cm$^{-1}$ will dissociate the DyCl$^+$ molecule.  

In order to further quantify this process we have used the RCI-VB method to determine
the  electronic transition dipole moments. We find that their values  range from $0.001 ea_0$ to $0.2 ea_0$ at the equilibrium separations of the $\Omega$ = 8 and 7 attractive potentials, where $e$ is the charge of the electron. Of the thirty-two repulsive $\Omega'$=6, 7, 8, and 9 potentials there are only fourteen  potentials with transition dipole moments larger than $0.1 ea_0$ that dissociate to the Dy$^{+}$+Cl limit. The seven excited potentials in the upper panel of Fig.~\ref{curves}  with a relatively large transition dipole moment to the $\Omega=8$ ground state are highlighted in red, whereas the other seven potentials with large dipole moments to the ground $\Omega$=7 state are highlighted in blue.

The selectivity of the transition dipole moment follows from the ground and
excited electronic wavefunctions.  The  wavefunction of the  X state is
dominated by the Dy$^{++}$Cl$^-$ configuration, with its single active
4f$^{10}$ Dy$^{++}$ shell with total core electron projection quantum number $\Omega_{\rm
c}=8$, and more importantly with a closed 3p$^6$ Cl valence shell and projection
quantum number $\Omega_{\rm v}=0^+$, such that $\Omega=\Omega_{\rm
c}+\Omega_{\rm v}$.  The electronic wavefunction for the optically-active repulsive
states is  controlled by interactions between the $^2$S valence electron of
the Dy$^+$ 6s  shell and the  $^2$P hole of  the Cl 2p$^5$ shell. An optical transition will excite an electron from 
the 3p$^6$ shell of Cl$^-$ into the 6s shell of Dy$^+$ without affecting the 4f$^{10}$($\Omega_c$=8) 
core. There are only a few one-electron transitions from the $\Omega$=8 ground state into  
repulsive excited states. In fact, selection rules suggest that only three $\Omega'$=7, one $\Omega'$=8, 
and three $\Omega'$=9 excited states have the same 4f$^{10}$ core state as the 
initial $\Omega$=8 ground state.  All other transitions are one or more orders of magnitude 
weaker, because they involve a change of state of two electrons. For the ground $\Omega$=7 potential 
a similar reasoning shows that only three $\Omega'$=6, one $\Omega'$=7, and three $\Omega'$=8 
excited states have significant transition dipole moments.

Using the procedures described in Ref.~\onlinecite{Chen2011} and a Hund's case (c) coupling scheme to describe the
molecular rotation, we predict the photodissociation cross section as a function of
laser frequency $\nu$, and compare the results with experimental data, as presented in section (\ref{ResultSec}).

\section{Experimental Methods}\label{ExptSec}

\begin{figure}[t]
 \centering
 \includegraphics[width=8.5cm]{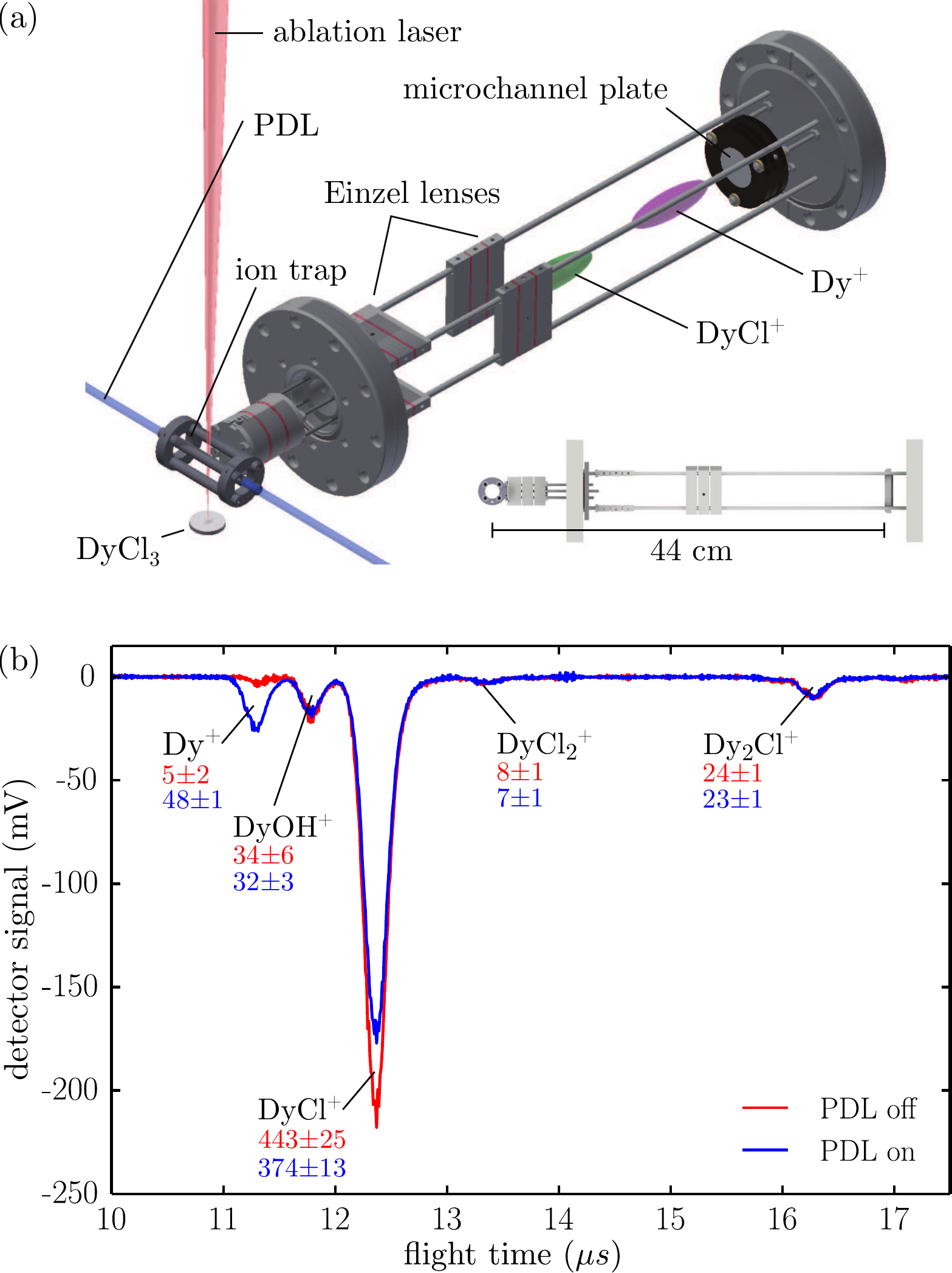}
 \caption{(a) A 3-D rendering of the ion trap and ToF-MS apparatus, with an artist's impression of species in the drift tube, separated due to their different masses. (b) Typical time-of-flight mass spectrometer traces used to determine the photodissociation yield, where dips represent arrival of ions at the detector. The measured number of ions in each peak is given underneath the species name, and the PDL wavenumber is $\sim45000$~cm$^{-1}$.}\label{ToFTrace}
\end{figure}

The DyCl$^+$ photodissociation spectrum was recorded using an integrated ion trap and time-of-flight mass spectrometer (ToF-MS) system, shown in Fig.~\ref{ToFTrace}(a), which has been described in previous work \cite{Schowalter2012, Puri2014}. Here we provide an overview of the experimenal setup and procedures, where particular attention is paid to details specific to this report, and refer the reader to the above Refs. for a more complete description of the apparatus.

A pressed, annealed pellet of DyCl$_3$, situated $\sim3$~cm away from the center of a linear quadrupole ion trap (LQT), is initially ablated by a single focused pulse from a 1064~nm Nd:YAG laser. The LQT, which has a field radius $r_0 = 7.92$~mm and an electrode radius $r_e = 3.18$~mm, operates at a frequency $\Omega_\textnormal{rf} = 2\pi\times 375$~kHz, and is switched on $\sim50~\mu$s after the ablation pulse. Non-linear motional resonances in the secular ion motion are exploited in order to reduce trapped species from the initial ablation yield other than DyCl$^+$, the most prominent of which is DyOH$^+$, as described in the following. The initial trapping amplitude $V_\textnormal{rf}=78$~V, which corresponds to a Matheiu-$q$ parameter for DyCl$^+$ (isotopic average mass $m = 197$~amu) of $q = 4eV_\textnormal{rf}/(m r_0^2\Omega_\textnormal{rf}^2) = 0.44$. At this $V_\textnormal{rf}$, the trap drives nonlinear resonant motion \cite{Eades1993} in Dy$^+$ ions, thus making them unstable (and therefore absent after several rf cycles) upon initial capture of the ablation products. After approximately $300$~ms, $V_\textnormal{rf}$ is ramped over $200$~ms to $85$~V, where a nonlinear motional resonance in DyOH$^+$ is driven. After 800~ms at this trapping amplitude, most of the DyOH$^+$ has escaped the trap, and only DyCl$^+$ ions and heavier (Dy$_2$Cl$^+$ is the only significant contributor) remain. The voltage is then ramped down over $200$~ms to $V_\textnormal{rf} = 65V$, where Dy$_2$Cl$^+$ is less prominent due to the lower mass-dependent trap depth $D_r \propto 1/m$, yet both Dy$^+$ and DyCl$^+$ (for which $q=0.364$) are stable. There remains a small background of Dy$^+$, DyOH$^+$ and Dy$_2$Cl$^+$ after mass filtering: an average Dy$^+$ background was measured, and incorporated into the resulting photodissociation yields; and tests show that DyOH$^+$ and Dy$_2$Cl$^+$ do not photodissociate at the photon energy range considered here.

Once the loading and mass filtering stages are complete, 1.7~s after ablation, the DyCl$^+$ is illuminated by laser pulses, propagating along the LQT axis, from a pulsed dye laser (PDL) operating with pulse duration $\tau_\mathrm{p}\sim10$~ns, pulse energy $E_\mathrm{p}\sim0.5$~mJ, and repetition rate $f_\mathrm{rep} = 10$~Hz, for $t = 1.3$~s.

The PDL beam is expanded by a telescope at the output of the laser, and subsequently passed through two sequential irises, each with an aperture of radius $r=1.5$~mm, to ensure uniform intensity distribution across the ion cloud, whose mean radius at the expected maximum translational temperature, 3400~K, is predicted to be $\sim1.2$~mm.

After the PDL pulses have been applied, the ions are radially ejected from the LQT into the ToF-MS. Ejection is achieved by switching off the trapping rf field and simultaneously applying 2~kV and 1.8~kV DC respectively, with $<1~\mu$s rise-time, to the LQT electrode pairs furthest from and closest to the entrance to a 44~cm drift tube. To maximise signal, focussing of the in-flight ions is performed using Einzel lenses; and the ions are detected upon arrival at a microchannel plate. Fig.~\ref{ToFTrace}(b) shows typical ToF-MS signals with and without the PDL (extraction occurs after the same trapping duration for both cases), where the dips represent arrival of ions. The ToF-MS Dy$^+$ arrival time is calibrated by trapping and ejecting the Dy$^+$ ablation yield from a $99.9~\%$ pure Dy ingot, and we fit the remaining arrival times observed from the ablation of DyCl$_3$ according to $t = A\sqrt{m} + t_0$, where $t_0$ is a time offset imposed by the pulsing electronics, and A is a constant determined by the pulse characteristics and flight-tube length. By minimising the $\chi^2$ statistic, we predict the peak at around 11.8~$\mu$s to be DyOH$^+$ and not DyO$^+$, although it is important to note that the resolution afforded by the ToF-MS for species at 300~K is insufficient to discriminate between these masses with complete certainty. To improve the ToF-MS resolution, one may employ laser cooling techniques to increase phase-space density, as described in Ref.~\onlinecite{Schneider2014}. As betrayed in Fig.~\ref{ToFTrace}(b), small amounts of DyOH$^+$, Dy$_2$Cl$^+$, and DyCl$_2^+$ are present in the ion trap even after mass filtering, yet, fortuitously, these species did not detectably photodissociate at the photon energies considered here. Furthermore, our data demonstrate that DyOH$^+$ is not formed due to reactions between the Dy$^+$ produced during photodissociation of DyCl$^+$ and any background of water (or oxygen if the species is in fact DyO$^+$) at a detectable level. Our DyCl$^+$ sample consists of all naturally-abundant isotopes of the molecule, yet we consider only $^{161}$Dy$^{35}$Cl$^+$ for calculation purposes, since the isotope shift of the photodissociation cross section is expected to be negligible on the photon energy scales considered here.\newline

\begin{figure}[t]
\includegraphics[width=8.5cm]{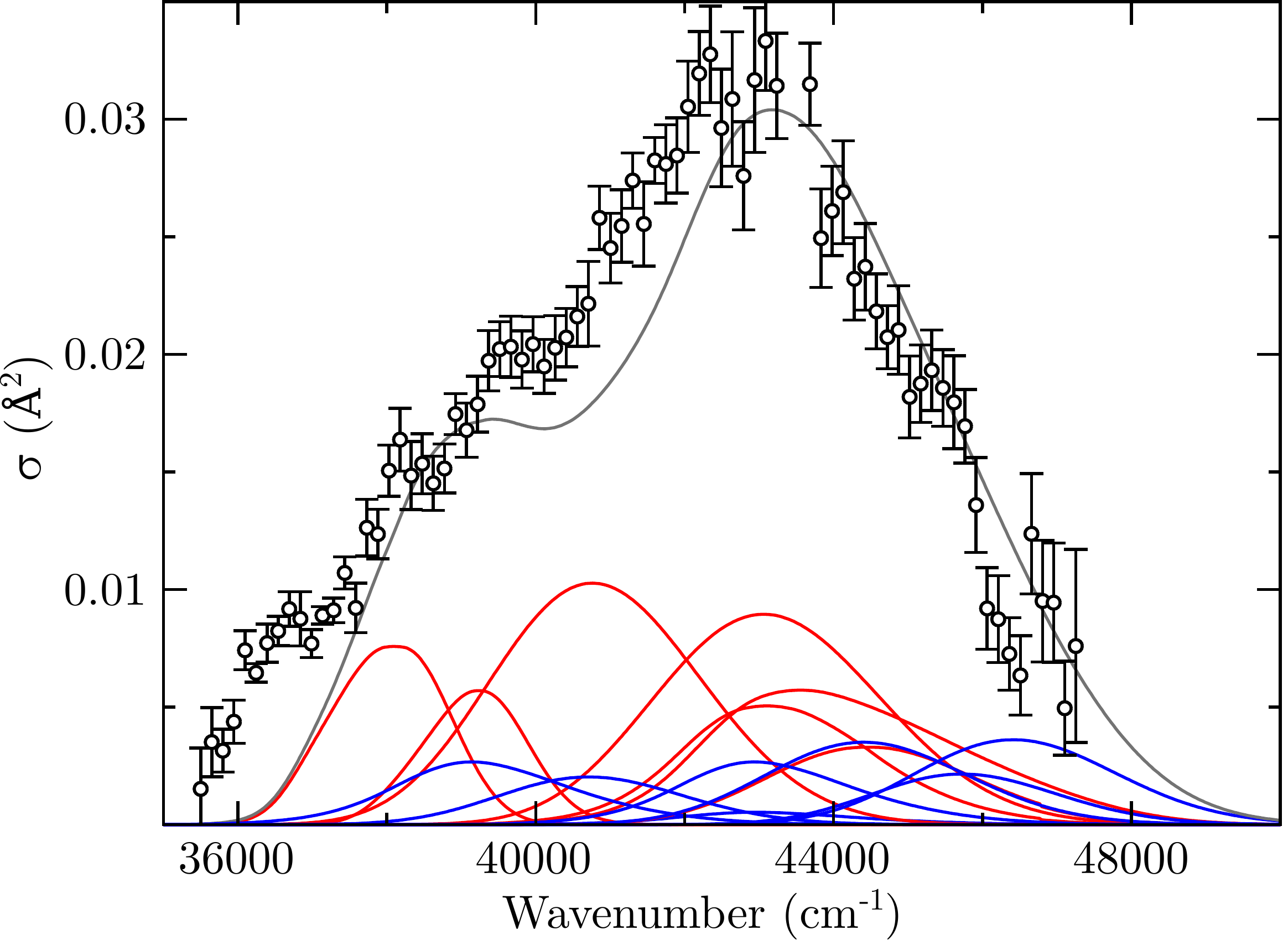}
\caption{(color online) The experimental (markers) and theoretical (grey line) 
thermalized photodissociation cross section of DyCl$^+$ molecular ions as a function of photon wavenumber $\nu/c$. The theoretical curves are for $^{161}$Dy$^{35}$Cl$^+$ at $T=300$~K.
The red and blue lines show partial contributions to the cross section due to transitions from vibrational levels of the $\Omega=8$ and 7
ground state potentials, respectively. These colors correspond to those used in Fig.~\ref{curves}.}
\label{pmd}
\end{figure}

From the ToF-MS traces, we infer the photodissociation cross section $\sigma$ at wavenumber $\nu$ according to
\begin{equation}
 \sigma = \frac{hc\nu}{\bar{I}t}\ln{\left(\frac{1}{1-\eta}\right)}
\end{equation}
where $h$ is Planck's constant, $c$ is the speed of light, $\bar{I} = f_\mathrm{rep}E_\mathrm{p}/\pi r^2$ is the time-averaged intensity of the PDL beam, and
\begin{equation}
\eta = \frac{N[\mathrm{Dy^+}]-N_{\mathrm{BG}}[\mathrm{Dy^+}]}{(N[\mathrm{Dy^+}]-N_{\mathrm{BG}}[\mathrm{Dy^+}])+N[\mathrm{DyCl^+}]}
\end{equation}
is the fractional photodissociation yield, in which $N[\mathrm{S}]$ is the measured number of species S arriving at the ion detector, and the BG subscript denotes the average background number, as measured with the PDL switched off.

A total of six different dye solutions including one exalite and five coumarin based dyes were used to span the range $35,500\leq\nu\leq47,400$~cm$^{-1}$ with intervals of $50$~cm$^{-1}$, where 10 measurements of $\eta$ were taken at each sampled $\nu$. The PDL pulse energy $E_\mathrm{p}$ was recorded for every pulse. Above $\nu \sim 46,000$~cm$^{-1}$, PDL output energies began to decrease, and attenuation of the pulses en-route to the LQT by both air and optics became increasingly apparent, leading to reduced signal-to-noise. Statistical error on the measurement of $\sigma$ arises due to shot-to-shot fluctuations in $E_\mathrm{p}$, along with variations in $\eta$.

\section{Results \& Discussion}\label{ResultSec}

Fig.~\ref{pmd} shows a comparison of our experimental and theoretical photodissociation cross sections. The markers represent the data, which are binned and averaged with bin widths of 150 cm$^{-1}$, and their error bars represent the standard error within each bin. The theoretical cross section has an unresolved double-peaked structure with a maximum around a photon energy of 43,000 cm$^{-1}$ and includes contributions from fourteen dominant transitions from the thermally populated $\Omega$=8 and 7 states and their corresponding $v$=0 and 1 levels. The maximum of the data approximately coincides with that of the predicted cross section. The data do not clearly reveal the predicted double-peaked sructure, yet they exhibit structure around 39,000 cm$^{-1}$ which might represent the predicted smaller peak. We find that the data lie mostly above the predicted curve at photon energies below 43,000 cm$^{-1}$, and above this value they agree well with the prediction. There appears to be structure around 36,500 cm$^{-1}$ present in the experiment but not in the theoretical prediction, which might point toward the existence of repulsive states whose energies are lower at the vertical transition turning point (illustrated by the dashed line in Figure~\ref{curves}) than those calculated here. Nevertheless, the agreement between experiment and theory is satisfactory given the accuracy of the potentials and dipole moments as well as the approximations made in evaluation of the cross section. It is important to note that no scaling was applied to parameters in the calculation for purposes of achieving a good match with the data.

Although calculations of molecular potentials have in this work been restricted to those corresponding to the four lowest-energy dissociation limits, we may speculate regarding the position of bound excited molecular
potentials in DyCl$^+$. For the purposes of laser cooling and quantum information experiments, we are particularly interested in the existence of electronic transitions which have optical frequencies and highly diagonal Franck-Condon factors (FCFs). For lanthanide-halides, it is generally expected that transitions involving the promotion of $4f$ or $6s$ electrons on the metal ion to $5d$ or $6p$ orbitals significantly affect the bonding characteristics of the molecule \cite{Gibson2003}, thus leading to FCFs which are non-diagonal in nature. On the other hand, intra-shell $f-f$ excitations within the Dy$^{++} [4f^{10}]$ metal core (see Ref.~\onlinecite{Spector1997} for free Dy$^{++}$ spectra) are not expected to affect the molecule's bonding characteristics, owing to the submerged nature of the $4f$ shell, as discussed in the context of lanthanide-trihalides in Ref.~\onlinecite{Lanza2008}. Such transitions are commonly exploited in Lanthanide-ion-doped solid-state laser materials, such as Nd$^{3+}[4f^3(^4F_{11/2} \leftarrow\, ^4I_{3/2})]$ in Nd$^{3+}$:YAG crystals, where vibrations and interactions with ligand fields lift the Laporte (parity) selection rule, leading to sharp spectral lines \cite{Hatanaka2014}. In free Dy$^{++}$, transitions within the $4f^{10}(^5I)$ core ground configuation, whose states should be slightly mixed, would occur in the near-infrared frequency range\cite{Spector1997}, with the largest interval being 11,014~cm$^{-1}$ for $4f^{10}(^5I_8 \leftarrow\, ^5I_4)$.  With parity mixing induced by the ligand field in Dy$^{++}$Cl$^-$, it is expected that these $f-f$ transitions will be strengthened. Should this be the case, we would expect a manifold of NIR-addressable, bound, excited potentials, whose equilibrium internuclear separations $r_e$ are close to that of the ground state. Such transitions, while narrow, would exhibit diagonal FCFs, thus potentially making them amenable to narrow-line laser cooling and coherent quantum manipulation. As such, they will form the basis for future work on DyCl$^+$.

\section{Summary}\label{SummarySec}

We have computed and measured the photodissociation cross section of DyCl$^+$ for photon energies between 35,500 cm$^{-1}$ and 47,500 cm$^{-1}$. Calculations show that there are 108 potential curves that dissociate to lowest spin-orbit ground states of
Dy$^+$ and Cl, which highlights the complex role played by the open $4f^{10}$ shell of Dy$^+$.
Only nine of these potentials are attractive and correspond to the ``ionically'' bonded Dy$^{++}$Cl$^-$,
where the outer 6s electron of Dy has transferred into the Cl $2p^5$ shell. 
All other potential curves corresponding to these dissociation limits are repulsive. Our experimental photodissociation cross section, obtained using an integrated ion trap and time-of-flight mass spectrometer, is found to agree satisfactorily with the calculations, given the myriad repulsive potentials and approximations made in the predicted transition moments. We observe a broad, asymmetric photodissociation cross section peaked at around 43,000 cm$^{-1}$, which is well-represented by calculations assuming thermally populated internal molecular states. Future work will aim to probe the existence of optically-addressable, vertical $f-f$ transitions in the molecule, towards its use in laser cooling, ultracold chemisty, and quantum information experiments.
 
\section{Acknowledgements}
AD acknowledges valuable discussions with Michael Heaven of Emory University. This work was supported by National Science Foundation (NSF, grant Nos. PHY-1205311, PHY-1308573), Army Research Office (MURI-ARO, grant Nos. W911NF-15-1-0121, W911NF-14-1-0378), and Air Force Office of Scientific Research (AFOSR, grant No. FA-14-1-0321) grants.


\begin{thebibliography}{34}%
\makeatletter
\providecommand \@ifxundefined [1]{%
 \@ifx{#1\undefined}
}%
\providecommand \@ifnum [1]{%
 \ifnum #1\expandafter \@firstoftwo
 \else \expandafter \@secondoftwo
 \fi
}%
\providecommand \@ifx [1]{%
 \ifx #1\expandafter \@firstoftwo
 \else \expandafter \@secondoftwo
 \fi
}%
\providecommand \natexlab [1]{#1}%
\providecommand \enquote  [1]{``#1''}%
\providecommand \bibnamefont  [1]{#1}%
\providecommand \bibfnamefont [1]{#1}%
\providecommand \citenamefont [1]{#1}%
\providecommand \href@noop [0]{\@secondoftwo}%
\providecommand \href [0]{\begingroup \@sanitize@url \@href}%
\providecommand \@href[1]{\@@startlink{#1}\@@href}%
\providecommand \@@href[1]{\endgroup#1\@@endlink}%
\providecommand \@sanitize@url [0]{\catcode `\\12\catcode `\$12\catcode
  `\&12\catcode `\#12\catcode `\^12\catcode `\_12\catcode `\%12\relax}%
\providecommand \@@startlink[1]{}%
\providecommand \@@endlink[0]{}%
\providecommand \url  [0]{\begingroup\@sanitize@url \@url }%
\providecommand \@url [1]{\endgroup\@href {#1}{\urlprefix }}%
\providecommand \urlprefix  [0]{URL }%
\providecommand \Eprint [0]{\href }%
\providecommand \doibase [0]{http://dx.doi.org/}%
\providecommand \selectlanguage [0]{\@gobble}%
\providecommand \bibinfo  [0]{\@secondoftwo}%
\providecommand \bibfield  [0]{\@secondoftwo}%
\providecommand \translation [1]{[#1]}%
\providecommand \BibitemOpen [0]{}%
\providecommand \bibitemStop [0]{}%
\providecommand \bibitemNoStop [0]{.\EOS\space}%
\providecommand \EOS [0]{\spacefactor3000\relax}%
\providecommand \BibitemShut  [1]{\csname bibitem#1\endcsname}%
\let\auto@bib@innerbib\@empty
\bibitem [{\citenamefont {Carr}\ \emph {et~al.}(2009)\citenamefont {Carr},
  \citenamefont {DeMille}, \citenamefont {Krems},\ and\ \citenamefont
  {Ye}}]{Carr2009}%
  \BibitemOpen
  \bibfield  {author} {\bibinfo {author} {\bibfnamefont {L.~D.}\ \bibnamefont
  {Carr}}, \bibinfo {author} {\bibfnamefont {D.}~\bibnamefont {DeMille}},
  \bibinfo {author} {\bibfnamefont {R.~V.}\ \bibnamefont {Krems}}, \ and\
  \bibinfo {author} {\bibfnamefont {J.}~\bibnamefont {Ye}},\ }\href {\doibase
  10.1088/1367-2630/11/5/055049} {\bibfield  {journal} {\bibinfo  {journal}
  {New Journal of Physics}\ }\textbf {\bibinfo {volume} {11}},\ \bibinfo
  {pages} {055049} (\bibinfo {year} {2009})}\BibitemShut {NoStop}%
\bibitem [{\citenamefont {Schuster}\ \emph {et~al.}(2011)\citenamefont
  {Schuster}, \citenamefont {Bishop}, \citenamefont {Chuang}, \citenamefont
  {DeMille},\ and\ \citenamefont {Schoelkopf}}]{Schuster2011}%
  \BibitemOpen
  \bibfield  {author} {\bibinfo {author} {\bibfnamefont {D.~I.}\ \bibnamefont
  {Schuster}}, \bibinfo {author} {\bibfnamefont {L.~S.}\ \bibnamefont
  {Bishop}}, \bibinfo {author} {\bibfnamefont {I.~L.}\ \bibnamefont {Chuang}},
  \bibinfo {author} {\bibfnamefont {D.}~\bibnamefont {DeMille}}, \ and\
  \bibinfo {author} {\bibfnamefont {R.~J.}\ \bibnamefont {Schoelkopf}},\ }\href
  {\doibase 10.1103/PhysRevA.83.012311} {\bibfield  {journal} {\bibinfo
  {journal} {Physical Review A}\ }\textbf {\bibinfo {volume} {83}},\ \bibinfo
  {pages} {012311} (\bibinfo {year} {2011})}\BibitemShut {NoStop}%
\bibitem [{\citenamefont {Pa\v{s}teka}\ \emph {et~al.}(2015)\citenamefont
  {Pa\v{s}teka}, \citenamefont {Borschevsky}, \citenamefont {Flambaum},\ and\
  \citenamefont {Schwerdtfeger}}]{Pasteka2015a}%
  \BibitemOpen
  \bibfield  {author} {\bibinfo {author} {\bibfnamefont {L.~F.}\ \bibnamefont
  {Pa\v{s}teka}}, \bibinfo {author} {\bibfnamefont {A.}~\bibnamefont
  {Borschevsky}}, \bibinfo {author} {\bibfnamefont {V.~V.}\ \bibnamefont
  {Flambaum}}, \ and\ \bibinfo {author} {\bibfnamefont {P.}~\bibnamefont
  {Schwerdtfeger}},\ }\href {\doibase 10.1103/PhysRevA.92.012103} {\bibfield
  {journal} {\bibinfo  {journal} {Physical Review A}\ }\textbf {\bibinfo
  {volume} {92}},\ \bibinfo {pages} {012103} (\bibinfo {year}
  {2015})}\BibitemShut {NoStop}%
\bibitem [{\citenamefont {Meyer}\ \emph {et~al.}(2006)\citenamefont {Meyer},
  \citenamefont {Bohn},\ and\ \citenamefont {Deskevich}}]{Meyer2006}%
  \BibitemOpen
  \bibfield  {author} {\bibinfo {author} {\bibfnamefont {E.}~\bibnamefont
  {Meyer}}, \bibinfo {author} {\bibfnamefont {J.}~\bibnamefont {Bohn}}, \ and\
  \bibinfo {author} {\bibfnamefont {M.}~\bibnamefont {Deskevich}},\ }\href
  {\doibase 10.1103/PhysRevA.73.062108} {\bibfield  {journal} {\bibinfo
  {journal} {Physical Review A}\ }\textbf {\bibinfo {volume} {73}},\ \bibinfo
  {pages} {062108} (\bibinfo {year} {2006})}\BibitemShut {NoStop}%
\bibitem [{\citenamefont {Leanhardt}\ \emph {et~al.}(2011)\citenamefont
  {Leanhardt}, \citenamefont {Bohn}, \citenamefont {Loh}, \citenamefont
  {Maletinsky}, \citenamefont {Meyer}, \citenamefont {Sinclair}, \citenamefont
  {Stutz},\ and\ \citenamefont {Cornell}}]{Leanhardt2011}%
  \BibitemOpen
  \bibfield  {author} {\bibinfo {author} {\bibfnamefont {A.}~\bibnamefont
  {Leanhardt}}, \bibinfo {author} {\bibfnamefont {J.}~\bibnamefont {Bohn}},
  \bibinfo {author} {\bibfnamefont {H.}~\bibnamefont {Loh}}, \bibinfo {author}
  {\bibfnamefont {P.}~\bibnamefont {Maletinsky}}, \bibinfo {author}
  {\bibfnamefont {E.}~\bibnamefont {Meyer}}, \bibinfo {author} {\bibfnamefont
  {L.}~\bibnamefont {Sinclair}}, \bibinfo {author} {\bibfnamefont
  {R.}~\bibnamefont {Stutz}}, \ and\ \bibinfo {author} {\bibfnamefont
  {E.}~\bibnamefont {Cornell}},\ }\href {\doibase 10.1016/j.jms.2011.06.007}
  {\bibfield  {journal} {\bibinfo  {journal} {Journal of Molecular
  Spectroscopy}\ }\textbf {\bibinfo {volume} {270}},\ \bibinfo {pages} {1}
  (\bibinfo {year} {2011})}\BibitemShut {NoStop}%
\bibitem [{\citenamefont {Yun}\ and\ \citenamefont {Nam}(2013)}]{Yun2013}%
  \BibitemOpen
  \bibfield  {author} {\bibinfo {author} {\bibfnamefont {S.~J.}\ \bibnamefont
  {Yun}}\ and\ \bibinfo {author} {\bibfnamefont {C.~H.}\ \bibnamefont {Nam}},\
  }\href {\doibase 10.1103/PhysRevA.87.040302} {\bibfield  {journal} {\bibinfo
  {journal} {Physical Review A - Atomic, Molecular, and Optical Physics}\
  }\textbf {\bibinfo {volume} {87}},\ \bibinfo {pages} {040302} (\bibinfo
  {year} {2013})}\BibitemShut {NoStop}%
\bibitem [{\citenamefont {M\o~lhave}\ and\ \citenamefont
  {Drewsen}(2000)}]{Molhave2000}%
  \BibitemOpen
  \bibfield  {author} {\bibinfo {author} {\bibfnamefont {K.}~\bibnamefont
  {M\o~lhave}}\ and\ \bibinfo {author} {\bibfnamefont {M.}~\bibnamefont
  {Drewsen}},\ }\href {\doibase 10.1103/PhysRevA.62.011401} {\bibfield
  {journal} {\bibinfo  {journal} {Physical Review A}\ }\textbf {\bibinfo
  {volume} {62}},\ \bibinfo {pages} {011401} (\bibinfo {year}
  {2000})}\BibitemShut {NoStop}%
\bibitem [{\citenamefont {Ostendorf}\ \emph {et~al.}(2006)\citenamefont
  {Ostendorf}, \citenamefont {Zhang}, \citenamefont {Wilson}, \citenamefont
  {Offenberg}, \citenamefont {Roth},\ and\ \citenamefont
  {Schiller}}]{Ostendorf2006}%
  \BibitemOpen
  \bibfield  {author} {\bibinfo {author} {\bibfnamefont {A.}~\bibnamefont
  {Ostendorf}}, \bibinfo {author} {\bibfnamefont {C.}~\bibnamefont {Zhang}},
  \bibinfo {author} {\bibfnamefont {M.}~\bibnamefont {Wilson}}, \bibinfo
  {author} {\bibfnamefont {D.}~\bibnamefont {Offenberg}}, \bibinfo {author}
  {\bibfnamefont {B.}~\bibnamefont {Roth}}, \ and\ \bibinfo {author}
  {\bibfnamefont {S.}~\bibnamefont {Schiller}},\ }\href {\doibase
  10.1103/PhysRevLett.97.243005} {\bibfield  {journal} {\bibinfo  {journal}
  {Physical Review Letters}\ }\textbf {\bibinfo {volume} {97}},\ \bibinfo
  {pages} {243005} (\bibinfo {year} {2006})}\BibitemShut {NoStop}%
\bibitem [{\citenamefont {Hansen}\ \emph {et~al.}(2014)\citenamefont {Hansen},
  \citenamefont {Versolato}, \citenamefont {Kłosowski}, \citenamefont
  {Kristensen}, \citenamefont {Gingell}, \citenamefont {Schwarz}, \citenamefont
  {Windberger}, \citenamefont {Ullrich}, \citenamefont {L\'{o}pez-Urrutia},\
  and\ \citenamefont {Drewsen}}]{Hansen2014}%
  \BibitemOpen
  \bibfield  {author} {\bibinfo {author} {\bibfnamefont {A.~K.}\ \bibnamefont
  {Hansen}}, \bibinfo {author} {\bibfnamefont {O.~O.}\ \bibnamefont
  {Versolato}}, \bibinfo {author} {\bibfnamefont {L.}~\bibnamefont
  {Kłosowski}}, \bibinfo {author} {\bibfnamefont {S.~B.}\ \bibnamefont
  {Kristensen}}, \bibinfo {author} {\bibfnamefont {A.}~\bibnamefont {Gingell}},
  \bibinfo {author} {\bibfnamefont {M.}~\bibnamefont {Schwarz}}, \bibinfo
  {author} {\bibfnamefont {A.}~\bibnamefont {Windberger}}, \bibinfo {author}
  {\bibfnamefont {J.}~\bibnamefont {Ullrich}}, \bibinfo {author} {\bibfnamefont
  {J.~R.~C.}\ \bibnamefont {L\'{o}pez-Urrutia}}, \ and\ \bibinfo {author}
  {\bibfnamefont {M.}~\bibnamefont {Drewsen}},\ }\href {\doibase
  10.1038/nature12996} {\bibfield  {journal} {\bibinfo  {journal} {Nature}\
  }\textbf {\bibinfo {volume} {508}},\ \bibinfo {pages} {76} (\bibinfo {year}
  {2014})}\BibitemShut {NoStop}%
\bibitem [{\citenamefont {Rellergert}\ \emph {et~al.}(2013)\citenamefont
  {Rellergert}, \citenamefont {Sullivan}, \citenamefont {Schowalter},
  \citenamefont {Kotochigova}, \citenamefont {Chen},\ and\ \citenamefont
  {Hudson}}]{Rellergert2013}%
  \BibitemOpen
  \bibfield  {author} {\bibinfo {author} {\bibfnamefont {W.~G.}\ \bibnamefont
  {Rellergert}}, \bibinfo {author} {\bibfnamefont {S.~T.}\ \bibnamefont
  {Sullivan}}, \bibinfo {author} {\bibfnamefont {S.~J.}\ \bibnamefont
  {Schowalter}}, \bibinfo {author} {\bibfnamefont {S.}~\bibnamefont
  {Kotochigova}}, \bibinfo {author} {\bibfnamefont {K.}~\bibnamefont {Chen}}, \
  and\ \bibinfo {author} {\bibfnamefont {E.~R.}\ \bibnamefont {Hudson}},\
  }\href {\doibase 10.1038/nature11937} {\bibfield  {journal} {\bibinfo
  {journal} {Nature}\ }\textbf {\bibinfo {volume} {495}},\ \bibinfo {pages}
  {490} (\bibinfo {year} {2013})}\BibitemShut {NoStop}%
\bibitem [{\citenamefont {Saykally}\ and\ \citenamefont
  {Woods}(1981)}]{Saykally1981}%
  \BibitemOpen
  \bibfield  {author} {\bibinfo {author} {\bibfnamefont {R.~J.}\ \bibnamefont
  {Saykally}}\ and\ \bibinfo {author} {\bibfnamefont {R.~C.}\ \bibnamefont
  {Woods}},\ }\href {\doibase 10.1146/annurev.pc.32.100181.002155} {\bibfield
  {journal} {\bibinfo  {journal} {Annual Review of Physical Chemistry}\
  }\textbf {\bibinfo {volume} {32}},\ \bibinfo {pages} {403} (\bibinfo {year}
  {1981})}\BibitemShut {NoStop}%
\bibitem [{\citenamefont {Shenyavskaya}\ and\ \citenamefont
  {Gurvich}(1980)}]{Shenyavskaya1980}%
  \BibitemOpen
  \bibfield  {author} {\bibinfo {author} {\bibfnamefont {E.}~\bibnamefont
  {Shenyavskaya}}\ and\ \bibinfo {author} {\bibfnamefont {L.}~\bibnamefont
  {Gurvich}},\ }\href {\doibase 10.1016/0022-2852(80)90335-5} {\bibfield
  {journal} {\bibinfo  {journal} {Journal of Molecular Spectroscopy}\ }\textbf
  {\bibinfo {volume} {81}},\ \bibinfo {pages} {152} (\bibinfo {year}
  {1980})}\BibitemShut {NoStop}%
\bibitem [{\citenamefont {Balfour}\ and\ \citenamefont
  {Lindgren}(1980)}]{Balfour1980}%
  \BibitemOpen
  \bibfield  {author} {\bibinfo {author} {\bibfnamefont {W.~J.}\ \bibnamefont
  {Balfour}}\ and\ \bibinfo {author} {\bibfnamefont {B.}~\bibnamefont
  {Lindgren}},\ }\href {\doibase 10.1088/0031-8949/22/1/005} {\bibfield
  {journal} {\bibinfo  {journal} {Physica Scripta}\ }\textbf {\bibinfo {volume}
  {22}},\ \bibinfo {pages} {36} (\bibinfo {year} {1980})}\BibitemShut {NoStop}%
\bibitem [{\citenamefont {Ramsay}\ and\ \citenamefont
  {Sarre}(1982)}]{Ramsay1982}%
  \BibitemOpen
  \bibfield  {author} {\bibinfo {author} {\bibfnamefont {D.~A.}\ \bibnamefont
  {Ramsay}}\ and\ \bibinfo {author} {\bibfnamefont {P.~J.}\ \bibnamefont
  {Sarre}},\ }\href {\doibase 10.1039/f29827801331} {\bibfield  {journal}
  {\bibinfo  {journal} {Journal of the Chemical Society, Faraday Transactions
  2}\ }\textbf {\bibinfo {volume} {78}},\ \bibinfo {pages} {1331} (\bibinfo
  {year} {1982})}\BibitemShut {NoStop}%
\bibitem [{\citenamefont {Merer}\ \emph {et~al.}(1984)\citenamefont {Merer},
  \citenamefont {Cheung},\ and\ \citenamefont {Taylor}}]{Merer1984}%
  \BibitemOpen
  \bibfield  {author} {\bibinfo {author} {\bibfnamefont {A.}~\bibnamefont
  {Merer}}, \bibinfo {author} {\bibfnamefont {A.-C.}\ \bibnamefont {Cheung}}, \
  and\ \bibinfo {author} {\bibfnamefont {A.}~\bibnamefont {Taylor}},\ }\href
  {\doibase 10.1016/0022-2852(84)90190-5} {\bibfield  {journal} {\bibinfo
  {journal} {Journal of Molecular Spectroscopy}\ }\textbf {\bibinfo {volume}
  {108}},\ \bibinfo {pages} {343} (\bibinfo {year} {1984})}\BibitemShut
  {NoStop}%
\bibitem [{\citenamefont {Chanda}\ \emph {et~al.}(1995)\citenamefont {Chanda},
  \citenamefont {Ho}, \citenamefont {Dalby},\ and\ \citenamefont
  {Ozier}}]{Chanda1995}%
  \BibitemOpen
  \bibfield  {author} {\bibinfo {author} {\bibfnamefont {A.}~\bibnamefont
  {Chanda}}, \bibinfo {author} {\bibfnamefont {W.~C.}\ \bibnamefont {Ho}},
  \bibinfo {author} {\bibfnamefont {F.~W.}\ \bibnamefont {Dalby}}, \ and\
  \bibinfo {author} {\bibfnamefont {I.}~\bibnamefont {Ozier}},\ }\href
  {\doibase 10.1063/1.468976} {\bibfield  {journal} {\bibinfo  {journal} {The
  Journal of Chemical Physics}\ }\textbf {\bibinfo {volume} {102}},\ \bibinfo
  {pages} {8725} (\bibinfo {year} {1995})}\BibitemShut {NoStop}%
\bibitem [{\citenamefont {W\"{u}est}\ and\ \citenamefont
  {Merkt}(2004)}]{Wuest2004}%
  \BibitemOpen
  \bibfield  {author} {\bibinfo {author} {\bibfnamefont {A.}~\bibnamefont
  {W\"{u}est}}\ and\ \bibinfo {author} {\bibfnamefont {F.}~\bibnamefont
  {Merkt}},\ }\href {\doibase 10.1063/1.1621618} {\bibfield  {journal}
  {\bibinfo  {journal} {The Journal of chemical physics}\ }\textbf {\bibinfo
  {volume} {120}},\ \bibinfo {pages} {638} (\bibinfo {year}
  {2004})}\BibitemShut {NoStop}%
\bibitem [{\citenamefont {Goncharov}\ and\ \citenamefont
  {Heaven}(2006)}]{Goncharov2006}%
  \BibitemOpen
  \bibfield  {author} {\bibinfo {author} {\bibfnamefont {V.}~\bibnamefont
  {Goncharov}}\ and\ \bibinfo {author} {\bibfnamefont {M.~C.}\ \bibnamefont
  {Heaven}},\ }\href {\doibase 10.1063/1.2167356} {\bibfield  {journal}
  {\bibinfo  {journal} {The Journal of chemical physics}\ }\textbf {\bibinfo
  {volume} {124}},\ \bibinfo {pages} {64312} (\bibinfo {year}
  {2006})}\BibitemShut {NoStop}%
\bibitem [{\citenamefont {Merritt}\ \emph {et~al.}(2009)\citenamefont
  {Merritt}, \citenamefont {Bondybey},\ and\ \citenamefont
  {Heaven}}]{Merritt2009}%
  \BibitemOpen
  \bibfield  {author} {\bibinfo {author} {\bibfnamefont {J.~M.}\ \bibnamefont
  {Merritt}}, \bibinfo {author} {\bibfnamefont {V.~E.}\ \bibnamefont
  {Bondybey}}, \ and\ \bibinfo {author} {\bibfnamefont {M.~C.}\ \bibnamefont
  {Heaven}},\ }\href {\doibase 10.1063/1.3098554} {\bibfield  {journal}
  {\bibinfo  {journal} {The Journal of chemical physics}\ }\textbf {\bibinfo
  {volume} {130}},\ \bibinfo {pages} {144503} (\bibinfo {year}
  {2009})}\BibitemShut {NoStop}%
\bibitem [{\citenamefont {Chen}\ \emph {et~al.}(2011)\citenamefont {Chen},
  \citenamefont {Schowalter}, \citenamefont {Kotochigova}, \citenamefont
  {Petrov}, \citenamefont {Rellergert}, \citenamefont {Sullivan},\ and\
  \citenamefont {Hudson}}]{Chen2011}%
  \BibitemOpen
  \bibfield  {author} {\bibinfo {author} {\bibfnamefont {K.}~\bibnamefont
  {Chen}}, \bibinfo {author} {\bibfnamefont {S.~J.}\ \bibnamefont
  {Schowalter}}, \bibinfo {author} {\bibfnamefont {S.}~\bibnamefont
  {Kotochigova}}, \bibinfo {author} {\bibfnamefont {A.}~\bibnamefont {Petrov}},
  \bibinfo {author} {\bibfnamefont {W.~G.}\ \bibnamefont {Rellergert}},
  \bibinfo {author} {\bibfnamefont {S.~T.}\ \bibnamefont {Sullivan}}, \ and\
  \bibinfo {author} {\bibfnamefont {E.~R.}\ \bibnamefont {Hudson}},\ }\href
  {\doibase 10.1103/PhysRevA.83.030501} {\bibfield  {journal} {\bibinfo
  {journal} {Physical Review A}\ }\textbf {\bibinfo {volume} {83}},\ \bibinfo
  {pages} {030501} (\bibinfo {year} {2011})}\BibitemShut {NoStop}%
\bibitem [{\citenamefont {Antonov}\ and\ \citenamefont
  {Heaven}(2013)}]{Antonov2013}%
  \BibitemOpen
  \bibfield  {author} {\bibinfo {author} {\bibfnamefont {I.~O.}\ \bibnamefont
  {Antonov}}\ and\ \bibinfo {author} {\bibfnamefont {M.~C.}\ \bibnamefont
  {Heaven}},\ }\href {\doibase 10.1021/jp312362e} {\bibfield  {journal}
  {\bibinfo  {journal} {The journal of physical chemistry. A}\ }\textbf
  {\bibinfo {volume} {117}},\ \bibinfo {pages} {9684} (\bibinfo {year}
  {2013})}\BibitemShut {NoStop}%
\bibitem [{\citenamefont {Lanza}\ \emph {et~al.}(2008)\citenamefont {Lanza},
  \citenamefont {Varga}, \citenamefont {Kolonits},\ and\ \citenamefont
  {Hargittai}}]{Lanza2008}%
  \BibitemOpen
  \bibfield  {author} {\bibinfo {author} {\bibfnamefont {G.}~\bibnamefont
  {Lanza}}, \bibinfo {author} {\bibfnamefont {Z.}~\bibnamefont {Varga}},
  \bibinfo {author} {\bibfnamefont {M.}~\bibnamefont {Kolonits}}, \ and\
  \bibinfo {author} {\bibfnamefont {M.}~\bibnamefont {Hargittai}},\ }\href
  {\doibase 10.1063/1.2828537} {\bibfield  {journal} {\bibinfo  {journal} {The
  Journal of chemical physics}\ }\textbf {\bibinfo {volume} {128}},\ \bibinfo
  {pages} {074301} (\bibinfo {year} {2008})}\BibitemShut {NoStop}%
\bibitem [{\citenamefont {Hatanaka}\ and\ \citenamefont
  {Yabushita}(2014)}]{Hatanaka2014}%
  \BibitemOpen
  \bibfield  {author} {\bibinfo {author} {\bibfnamefont {M.}~\bibnamefont
  {Hatanaka}}\ and\ \bibinfo {author} {\bibfnamefont {S.}~\bibnamefont
  {Yabushita}},\ }\href {\doibase 10.1007/s00214-014-1517-2} {\bibfield
  {journal} {\bibinfo  {journal} {Theoretical Chemistry Accounts}\ }\textbf
  {\bibinfo {volume} {133}},\ \bibinfo {pages} {1517} (\bibinfo {year}
  {2014})}\BibitemShut {NoStop}%
\bibitem [{\citenamefont {Kotochigova}\ and\ \citenamefont
  {Tiesinga}(2005)}]{Kotochigova2005}%
  \BibitemOpen
  \bibfield  {author} {\bibinfo {author} {\bibfnamefont {S.}~\bibnamefont
  {Kotochigova}}\ and\ \bibinfo {author} {\bibfnamefont {E.}~\bibnamefont
  {Tiesinga}},\ }\href {\doibase 10.1063/1.2107607} {\bibfield  {journal}
  {\bibinfo  {journal} {The Journal of chemical physics}\ }\textbf {\bibinfo
  {volume} {123}},\ \bibinfo {pages} {174304} (\bibinfo {year}
  {2005})}\BibitemShut {NoStop}%
\bibitem [{\citenamefont {Lim}\ \emph {et~al.}(2005)\citenamefont {Lim},
  \citenamefont {Schwerdtfeger}, \citenamefont {Metz},\ and\ \citenamefont
  {Stoll}}]{Lim2005}%
  \BibitemOpen
  \bibfield  {author} {\bibinfo {author} {\bibfnamefont {I.~S.}\ \bibnamefont
  {Lim}}, \bibinfo {author} {\bibfnamefont {P.}~\bibnamefont {Schwerdtfeger}},
  \bibinfo {author} {\bibfnamefont {B.}~\bibnamefont {Metz}}, \ and\ \bibinfo
  {author} {\bibfnamefont {H.}~\bibnamefont {Stoll}},\ }\href {\doibase
  10.1063/1.1856451} {\bibfield  {journal} {\bibinfo  {journal} {The Journal of
  chemical physics}\ }\textbf {\bibinfo {volume} {122}},\ \bibinfo {pages}
  {104103} (\bibinfo {year} {2005})}\BibitemShut {NoStop}%
\bibitem [{\citenamefont {Weigend}\ and\ \citenamefont
  {Ahlrichs}(2005)}]{Weigend2005}%
  \BibitemOpen
  \bibfield  {author} {\bibinfo {author} {\bibfnamefont {F.}~\bibnamefont
  {Weigend}}\ and\ \bibinfo {author} {\bibfnamefont {R.}~\bibnamefont
  {Ahlrichs}},\ }\href {\doibase 10.1039/b508541a} {\bibfield  {journal}
  {\bibinfo  {journal} {Physical chemistry chemical physics : PCCP}\ }\textbf
  {\bibinfo {volume} {7}},\ \bibinfo {pages} {3297} (\bibinfo {year}
  {2005})}\BibitemShut {NoStop}%
\bibitem [{\citenamefont {Puri}\ \emph {et~al.}(2014)\citenamefont {Puri},
  \citenamefont {Schowalter}, \citenamefont {Kotochigova}, \citenamefont
  {Petrov},\ and\ \citenamefont {Hudson}}]{Puri2014}%
  \BibitemOpen
  \bibfield  {author} {\bibinfo {author} {\bibfnamefont {P.}~\bibnamefont
  {Puri}}, \bibinfo {author} {\bibfnamefont {S.~J.}\ \bibnamefont
  {Schowalter}}, \bibinfo {author} {\bibfnamefont {S.}~\bibnamefont
  {Kotochigova}}, \bibinfo {author} {\bibfnamefont {A.}~\bibnamefont {Petrov}},
  \ and\ \bibinfo {author} {\bibfnamefont {E.~R.}\ \bibnamefont {Hudson}},\
  }\href {\doibase 10.1063/1.4885363} {\bibfield  {journal} {\bibinfo
  {journal} {The Journal of chemical physics}\ }\textbf {\bibinfo {volume}
  {141}},\ \bibinfo {pages} {014309} (\bibinfo {year} {2014})}\BibitemShut
  {NoStop}%
\bibitem [{\citenamefont {Herzberg}(1957)}]{Herzberg1957}%
  \BibitemOpen
  \bibfield  {author} {\bibinfo {author} {\bibfnamefont {G.}~\bibnamefont
  {Herzberg}},\ }\href@noop {} {\emph {\bibinfo {title} {{Molecular Spectra and
  Molecular Structure}}}}\ (\bibinfo  {publisher} {D. van Nostrand company,
  Inc},\ \bibinfo {year} {1957})\BibitemShut {NoStop}%
\bibitem [{\citenamefont {Lefebvre-Brion}\ and\ \citenamefont
  {Field}(2004)}]{Lefebvre-Brion2004}%
  \BibitemOpen
  \bibfield  {author} {\bibinfo {author} {\bibfnamefont {H.}~\bibnamefont
  {Lefebvre-Brion}}\ and\ \bibinfo {author} {\bibfnamefont {R.~W.}\
  \bibnamefont {Field}},\ }\href {\doibase 10.1016/B978-012441455-6/50010-X}
  {\emph {\bibinfo {title} {The Spectra and Dynamics of Diatomic Molecules}}}\
  (\bibinfo  {publisher} {Elsevier},\ \bibinfo {year} {2004})\ pp.\ \bibinfo
  {pages} {469--549}\BibitemShut {NoStop}%
\bibitem [{\citenamefont {Schowalter}\ \emph {et~al.}(2012)\citenamefont
  {Schowalter}, \citenamefont {Chen}, \citenamefont {Rellergert}, \citenamefont
  {Sullivan},\ and\ \citenamefont {Hudson}}]{Schowalter2012}%
  \BibitemOpen
  \bibfield  {author} {\bibinfo {author} {\bibfnamefont {S.~J.}\ \bibnamefont
  {Schowalter}}, \bibinfo {author} {\bibfnamefont {K.}~\bibnamefont {Chen}},
  \bibinfo {author} {\bibfnamefont {W.~G.}\ \bibnamefont {Rellergert}},
  \bibinfo {author} {\bibfnamefont {S.~T.}\ \bibnamefont {Sullivan}}, \ and\
  \bibinfo {author} {\bibfnamefont {E.~R.}\ \bibnamefont {Hudson}},\ }\href
  {\doibase 10.1063/1.3700216} {\bibfield  {journal} {\bibinfo  {journal} {The
  Review of scientific instruments}\ }\textbf {\bibinfo {volume} {83}},\
  \bibinfo {pages} {043103} (\bibinfo {year} {2012})}\BibitemShut {NoStop}%
\bibitem [{\citenamefont {Eades}\ \emph {et~al.}(1993)\citenamefont {Eades},
  \citenamefont {Johnson},\ and\ \citenamefont {Yost}}]{Eades1993}%
  \BibitemOpen
  \bibfield  {author} {\bibinfo {author} {\bibfnamefont {D.~M.}\ \bibnamefont
  {Eades}}, \bibinfo {author} {\bibfnamefont {J.~V.}\ \bibnamefont {Johnson}},
  \ and\ \bibinfo {author} {\bibfnamefont {R.~A.}\ \bibnamefont {Yost}},\
  }\href {\doibase 10.1016/1044-0305(93)80017-S} {\bibfield  {journal}
  {\bibinfo  {journal} {Journal of the American Society for Mass Spectrometry}\
  }\textbf {\bibinfo {volume} {4}},\ \bibinfo {pages} {917} (\bibinfo {year}
  {1993})}\BibitemShut {NoStop}%
\bibitem [{\citenamefont {Schneider}\ \emph {et~al.}(2014)\citenamefont
  {Schneider}, \citenamefont {Schowalter}, \citenamefont {Chen}, \citenamefont
  {Sullivan},\ and\ \citenamefont {Hudson}}]{Schneider2014}%
  \BibitemOpen
  \bibfield  {author} {\bibinfo {author} {\bibfnamefont {C.}~\bibnamefont
  {Schneider}}, \bibinfo {author} {\bibfnamefont {S.~J.}\ \bibnamefont
  {Schowalter}}, \bibinfo {author} {\bibfnamefont {K.}~\bibnamefont {Chen}},
  \bibinfo {author} {\bibfnamefont {S.~T.}\ \bibnamefont {Sullivan}}, \ and\
  \bibinfo {author} {\bibfnamefont {E.~R.}\ \bibnamefont {Hudson}},\ }\href
  {\doibase 10.1103/PhysRevApplied.2.034013} {\bibfield  {journal} {\bibinfo
  {journal} {Physical Review Applied}\ }\textbf {\bibinfo {volume} {2}},\
  \bibinfo {pages} {034013} (\bibinfo {year} {2014})}\BibitemShut {NoStop}%
\bibitem [{\citenamefont {Gibson}(2003)}]{Gibson2003}%
  \BibitemOpen
  \bibfield  {author} {\bibinfo {author} {\bibfnamefont {J.~K.}\ \bibnamefont
  {Gibson}},\ }\href {\doibase 10.1021/jp035003n} {\bibfield  {journal}
  {\bibinfo  {journal} {The Journal of Physical Chemistry A}\ }\textbf
  {\bibinfo {volume} {107}},\ \bibinfo {pages} {7891} (\bibinfo {year}
  {2003})}\BibitemShut {NoStop}%
\bibitem [{\citenamefont {Spector}\ \emph {et~al.}(1997)\citenamefont
  {Spector}, \citenamefont {Sugar},\ and\ \citenamefont {Wyart}}]{Spector1997}%
  \BibitemOpen
  \bibfield  {author} {\bibinfo {author} {\bibfnamefont {N.}~\bibnamefont
  {Spector}}, \bibinfo {author} {\bibfnamefont {J.}~\bibnamefont {Sugar}}, \
  and\ \bibinfo {author} {\bibfnamefont {J.-F.}\ \bibnamefont {Wyart}},\ }\href
  {\doibase 10.1364/JOSAB.14.000511} {\bibfield  {journal} {\bibinfo  {journal}
  {Journal of the Optical Society of America B}\ }\textbf {\bibinfo {volume}
  {14}},\ \bibinfo {pages} {511} (\bibinfo {year} {1997})}\BibitemShut
  {NoStop}%
\end{thebibliography}
\end{document}